\documentclass[12pt]{article}

\usepackage[dvips]{graphicx}
\usepackage{amsmath}
\usepackage{amssymb}
\usepackage{bbm}
\DeclareSymbolFont{AMSb}{U}{msb}{m}{n}
\DeclareSymbolFontAlphabet{\Bbb}{AMSb}
\newcommand{\N}{\mathbbm{N}}
\newcommand{\Z}{\mathbbm{Z}}

\newcommand{\T}{\mathbbm{T}}
\newcommand{\e}{\mathrm{e}}
\newcommand{\alg}{\mathcal{A}}
\newcommand{\qsc}{\mathcal{C}}
\newcommand{\hil}{\mathcal{H}}
\newcommand{\matr}{\mathcal{M}}
\newcommand{\Tr}{\mathrm{Tr}\,}
\newcommand{\idty}{\mathbbm{1}}

\title{Entropy growth of shift-invariant states on a quantum spin chain}
\author{
M.~Fannes$^{\,1,\,}$\footnote{E-mail:
mark.fannes@fys.kuleuven.ac.be}, 
B.~Haegeman$^{\,1,\,}$\footnote{E-mail:
bart.haegeman@fys.kuleuven.ac.be},
M.~Mosonyi$^{\,2,\,}$\footnote{E-mail:
mosonyi@chardonnay.math.bme.hu}
}

\begin{document}

\maketitle

\begin{center}
$^{\,1}$ Instituut voor Theoretische Fysica \\
K.U.Leuven \\
Celestijnenlaan~200D, B-3001 Heverlee, Belgium \\[6pt]
$^{\,2}$ Mathematical Institute \\ 
Budapest University of Technology and Economics \\
H-1521 Budapest XI. Sztoczek u. 2, Hungary\\

\end{center}

\begin{abstract}
\noindent 
We study the entropy of pure shift-invariant states on a  quantum
spin chain. Unlike the classical case, the local restrictions to
intervals of length $N$ are typically mixed and have therefore a 
non-zero entropy $S_N$ which is, moreover, monotonically increasing
in $N$. We are interested in the asymptotics of the total entropy. We
investigate in detail a class of states derived from quasi-free
states on a CAR~algebra. These are characterised by a measurable
subset of the unit interval. As the entropy density is known to
vanishes, $S_N$ is sublinear in $N$. For states corresponding to
unions of finitely many intervals, $S_N$ is shown to grow slower than
$(\log N)^2$. Numerical calculations suggest a $\log N$ behaviour.
For the case with infinitely many intervals, we present a class of
states for which the entropy $S_N$ increases as $N^\alpha$ where
$\alpha$ can take any value in $(0,1)$.
\end{abstract}

\section{Introduction}

In quantum statistical mechanics, one-dimensional lattice systems,
the so-called spin chains, are far from fully understood. One of the
obstacles for a systematic study is the complicated correlations that
can occur. This is even possible for pure states, which are trivial
for classical spin chains. Due to these quantum correlations, it is
often very hard to explicitly specify a state. Only a few classes can
be studied in detail, including the  product states, the finitely
correlated states~\cite{fnw} and the states derived from quasi-free
states on the CAR~algebra~\cite{br,ek}.

Let us denote by $\rho_N$ the density matrix  of the restriction of a
translation-invariant state $\rho$ on a spin chain to $N$ consecutive
spins. The von~Neumann entropy $S_N:=S(\rho_N)$ has proved to be a
very useful quantity in the study of quantum correlations. For
ergodic translation-invariant states, $\rho_N$ is essentially
concentrated on a subspace of dimension $\exp(N s(\rho))$~\cite{hp}.
Here, $s(\rho)$ is the entropy density of $\rho$. The compression of
$\rho_N$ from the full dimension $d^N$ of $N$ spins to $\exp(N
s(\rho))$ lies e.g.\ at the basis of DMRG~computations~\cite{pwkh}.
One may conjecture that $s(\rho)=0$ for pure states $\rho$, which
should allow for a very efficient compression. For pure states, $S_N$
is also the unique reasonable measure for the entanglement of this
interval with the rest of the chain~\cite{nc} and it measures therefore
the resources of the state for quantum computing purposes. 

For pure product states $S_N$ vanishes for all $N$, this is in fact 
completely analogous to the classical spin chain. For pure finitely
correlated states $S_N$ is uniformly bounded, a behaviour that is
certainly not expected to be generic.  

In this paper, we study the entropy $S_N$ for translation-invariant
pure states derived from quasi-free states on the CAR~algebra. Here,
the entropy density is known to vanish and we investigate the
sublinear growth of the entropy $S_N$ when $N\to\infty$. We show that
$S_N$ increases much faster with $N$ than in the previous cases. For
the simplest quasi-free states, the entropy behaves as $\log N$. 
We shall also present a more involved example for which the
entropy increases as $N^\alpha$ with $\alpha$ arbitrarily close to 1.

The construction of pure shift-invariant quasi-free states is
recalled in Section~\ref{sec:quafree}. Such states are characterised
by a subset of the unit interval. In Section~\ref{sec:asympt} we
prove that the asymptotics of $S_N$ as $N\to\infty$ can be obtained
by a quadratic approximation of the entropy. The entropy growth of
quasi-free states given by a set consisting of finitely many
intervals is studied in Section~\ref{sec:finite}. Finally,
Section~\ref{sec:infinite} is devoted to the infinitely many
intervals case.

\section{Quasi-free states on the spin chain} \label{sec:quafree}

In this section we show, following~\cite{ek}, how a quasi-free state
on the CAR~algebra can be used to define a state on the spin chain
algebra. After the introductory definitions, we explain how both
algebras can be retrieved as subalgebras of a larger algebra. This
construction permits to transfer translation-invariant states
from the CAR~algebra to the spin chain algebra. This idea is then
applied to quasi-free states.

\subsection{CAR algebra and spin chain algebra}

Let $\hil$ be the Hilbert space $\ell^2(\Z)$, in which  $\{\delta_k
\,:\, k\in\Z\}$ forms an orthonormal basis, where $\delta_k$ is the
characteristic function of the integer number $k$. Let $\alg$ be the
CAR~algebra corresponding to $\hil$. It is the C*-algebra generated
by $\idty$ and $\{c_k \,:\, k\in\Z\}$, satisfying the canonical
anticommutation relations
\begin{equation*}
 c_k c_l=-c_l c_k \qquad c_k^* c_l=\delta_{k,l}\idty-c_l c_k^*.
\end{equation*}

The parity automorphism $\alpha$ on $\alg$ is defined by
$\alpha(c_k)=-c_k$. Let $\alg_+$ be the fixed point algebra of
$\alpha$, i.e.\ $\alg_+ = \{a\in\alg \,:\, \alpha(a)=a\}$. The
elements of $\alg_+$ are called even, while those of $\alg_- :=
\{a\in\alg \,:\, \alpha(a)=-a\}$ are odd. Obviously, $\alg =
\alg_++\alg_-$. The shift auto\-mor\-phism $\gamma$ is defined by
$\gamma(c_k) = c_{k+1}$.

The quantum spin chain is the UHF~algebra
\begin{equation*}
 \qsc := \bigotimes_{k=-\infty}^{+\infty} \matr_2,
\end{equation*}
where $\matr_2$ is the algebra of $2\times 2$ matrices.
Let $e_{11}^k$, $e_{12}^k$, $e_{21}^k$ and $e_{22}^k$ denote
the standard matrix units of $\matr_2$ embedded into the $k$th
factor of $\qsc$. The following relations hold:
\begin{align} 
 e_{ab}^k e_{cd}^l 
 &= e_{cd}^l e_{ab}^k \quad \text{when $k\ne l$,} 
\nonumber \\
 e_{ab}^k e_{cd}^k 
 &= \delta_{b,c} e_{ad}^k,
\nonumber \\
 \left(e_{ab}^k\right)^* 
 &= e_{ba}^k,
\label{eq:matrunit} \\
 e_{11}^k+e_{22}^k 
 &= \idty.
\nonumber 
\end{align}
Any algebra generated by elements $\{E_{ab}^k \,:\, a,b\in\{1,2\},\
k\in\Z\}$ satisfying the above relations is isomorphic to $\qsc$.

\subsection{Jordan-Wigner isomorphism}

Let $\alg_n$ be the algebra generated by $\{c_k \,:\, 0\leq k\leq
n-1\}$ and let $\qsc_n=\bigotimes_{k=0}^{n-1} \matr_2$. It is
well-known that $\alg_n$ is isomorphic to $\qsc_n$ for all $n\in\N$.
An explicit isomorphism is given by the so-called  Jordan-Wigner
isomorphism given in terms of matrix units in $\alg_n$ by
\begin{equation*}
 E_{11}^k:=c_k^*c_k, \quad
 E_{22}^k:=c_k c_k^*, \quad
 E_{12}^k:=A_k c_k^*, \quad
 E_{21}^k:=A_k c_k. \quad
\end{equation*}
Here we introduced
\begin{equation*}
 \sigma_k^z:=2c_k^*c_k-\idty, \qquad A_k:=\prod_{l=0}^{k-1}\sigma_l^z.
\end{equation*}
The set $\{E_{ab}^k \,:\, a,b\in\{1,2\},0\leq k\leq n-1\}$ generates
$\alg_n$ and the operators $E_{ab}^k$ satisfy the same 
relations~(\ref{eq:matrunit}) as the matrix units $e_{ab}^k$ of
$C_n$.

A first idea would be to extend this isomorphism to an isomorphism
from $\alg$ to $\qsc$. However, it is impossible to extend this
definition to negative $k$'s in such a way that the isomorphism
intertwines the shifts in $\alg$ and $\qsc$. This property is needed
to transport translation-invariance from $\alg$ to $\qsc$.

One way to circumvent this problem is the following. Enlarge $\alg$
to $\hat\alg$ by adding a new element $T$ that has the following
properties
\begin{align*} 
 &T^*=T,\quad T^2=\idty\quad \text{(i.e.\ $T$ is a self-adjoint
 unitary)} \\
 &Tc_k T = \begin{cases} c_k &\text{if $k\ge 0$}\\ -c_k &\text{if
 $k<0$.} \end{cases}
\end{align*}    
Any element of $\hat\alg$ can uniquely be written in the form $a+Tb$
with $a$ and $b$ from $\alg$. Therefore, $\hat\alg=\alg+T\alg$. Note
that formally $T=\prod_{k=-1}^{-\infty}\sigma_k^z$.

A state $\varphi$ on $\alg$ can be extended to a state $\hat\varphi$
on $\hat\alg$ by $\hat\varphi(a+Tb):=\varphi(a)$ and the extensions of
the automorphisms $\alpha$ and $\gamma$ are
\begin{equation*}
 \hat\alpha(a+Tb) := \alpha(a)+T\alpha(b) 
 \qquad\text{and}\qquad
 \hat\gamma(a+Tb) := \gamma(a)+T\sigma_0^z\gamma(b).
\end{equation*}
We define another automorphism $\beta$ on $\hat\alg$ by
$\beta(a+Tb) := a-Tb$. The fixed point algebra of
$\beta^{-1}\hat\alpha$ will be denoted by $\check\alg$, i.e.,
\begin{align*}
 \check\alg
 &= \{ a+Tb \in \hat\alg \,:\, \hat\alpha(a+Tb) = \beta(a+Tb) \} \\
 &= \{ a+Tb \in \hat\alg \,:\, \alpha(a)=a,\, \alpha(b)=-b \} \\
 &= \alg_+ + T \alg_-.
\end{align*}
The restriction of a state $\hat\varphi$ on $\hat\alg$ to a state on
$\check\alg$ will be denoted by $\check\varphi$. Because the
automorphisms $\hat\alpha$ and $\hat\gamma$ leave the subalgebra
$\check\alg$ invariant, they can be restricted to $\check\alg$.
Denote these restrictions by $\check\alpha$ and $\check\gamma$.

Let $\varphi$ be an even state, i.e., it vanishes on odd elements or,
equivalently, $\varphi\circ\alpha=\varphi$. It is easy to see that
also $\check\varphi\circ\check\alpha=\check\varphi$, thus
$\check\varphi$ is an even state on $\check\alg$. Similarly, let
$\varphi$ be a translation-invariant state on $\alg$, i.e.\ 
$\varphi\circ\gamma=\varphi$, then $\check\varphi \circ \check\gamma
= \check\varphi$, thus $\check\varphi$ is a translation-invariant
state on $\check\alg$.

Now, define
\begin{equation*}
 \tilde E_{11}^k:=c_k^*c_k, \quad
 \tilde E_{22}^k:=c_k c_k^*, \quad
 \tilde E_{12}^k:=T A_k c_k^*, \quad
 \tilde E_{21}^k:=T A_k c_k. \quad
\end{equation*}
with
\begin{equation*}
 \sigma_k^z:=2c_k^*c_k-\idty, \qquad
 A_k := \begin{cases} \prod_{\ell=0}^{k-1} \sigma_\ell^z &\text{if } k>0 \\
       \idty &\text{if } k=0 \\ \prod_{\ell=k}^{-1} \sigma_\ell^z
       &\text{if } k<0. \end{cases}
\end{equation*}
One verifies that these operators satisfy the same commutation
relations as the matrix units of $\qsc$. Moreover,
$\check\gamma(\tilde E_{ab}^k)=\tilde E_{ab}^{k+1}$.

To summarise, we constructed an algebra $\hat\alg$ which contains
both $\alg$ and $\check\alg$ as subalgebras. This embedding is
compatible with the translations on the subalgebras. Moreover, we
established an isomorphism between $\check\alg$ and $\qsc$ which is
also compatible with the translations. This allows us to transfer
translation-invariant states from $\alg$ to $\qsc$.

Let $\varphi$ be a translation-invariant state. Such a state is
automatically even and is completely determined by the sequence
$\left(\varphi_n\right)_{n=0}^{+\infty}$,
where $\varphi_n$ is the restriction of $\varphi$ to $\alg_n$.
The density matrix $[\varphi_n]$ of $\varphi_n$ has entries
\begin{equation*}
 [\varphi_n]_{\underline j,\underline i} = 
 \varphi  \left( \prod_{k=0}^{n-1} E_{i_k j_k}^k \right),
 \qquad \underline i,\underline j\in \{1,2\}^n.
\end{equation*} 
The transferred state $\check\varphi$ is also translation-invariant
and so completely determined by its restriction to the subalgebras
$\{\qsc_n \,:\, n\in\N\}$ with density matrices
\begin{equation*}
 [\check\varphi_n]_{\underline j,\underline i} = \hat{\varphi}
 \left(\prod_{k=0}^{n-1} \tilde E_{i_k j_k}^k\right)
 \qquad \underline i,\underline j\in \{1,2\}^n.
\end{equation*} 
The expressions $\prod_{k=0}^{n-1} E_{i_k j_k}^k$ and
$\prod_{k=0}^{n-1} \tilde E_{i_k j_k}^k$ are both either odd or even.
When odd, $[\varphi_n]_{\underline j,\underline i}=
[\check\varphi_n]_{\underline j,\underline i}=0$, while when even,
since $Tc_k=c_k T$ for $k\ge 0$ and $T^2=\idty$, we get that
$\prod_{k=0}^{n-1}\tilde E_{i_k j_k}^k = \prod_{k=0}^{n-1} E_{i_k
j_k}^k$ and so $[\varphi_n]_{\underline i,\underline j} =
\left[\check\varphi_n\right]_{\underline i,\underline j}$. From this,
we conclude that the states $\varphi$ and $\check\varphi$ have the
same reduced density matrices. It follows immediately that if
$\varphi$ is pure, then also $\check\varphi$ is pure.

\subsection{Quasi-free states}

We apply the construction of the previous section to quasi-free
states on the CAR~algebra $\alg$. For these states an explicit
formula is known for the entropy of the restricted density matrices.
Because the corresponding states on $\qsc$ have the same restricted
density matrices, the same explicit formulas are available, as we
shall use in the following sections. The proofs of the theorems
mentioned in this subsection can be found in~\cite{af}. 

Let $\varphi$ be a quasi-free, gauge-invariant state on $\alg$,
i.e., $\varphi$ is given by the rule
\begin{equation*}
 \varphi(c_{i_1}^*\dots c_{i_m}^* c_{j_n}\dots c_{j_1}) =
 \delta_{m,n} \det\left( \left[ Q_{i_kj_l} \right]_{k,l=1}^n \right),
\end{equation*} 
where $Q$ is an operator on $\hil$, $0\leq Q\leq\idty$ and $Q_{ij} =
\langle \delta_i,Q\delta_j \rangle$ are the matrix elements of $Q$ in
the standard basis of $\hil$. The operator $Q$ is called the symbol
of the state $\varphi$. Obviously, $\varphi$ is even.

The quasi-free state $\varphi$ is translation-invariant if and only
if its symbol $Q$ is a Toeplitz operator, i.e., there exists a
sequence $\{q_k \,:\, k\in\Z\}$ such that $Q_{lk} = q(k-l)$. By the
Fourier transform,
\begin{equation} \label{eq:Qcomp}
 q^\land(\theta) = \sum_{k\in\Z} q(k) \mathrm e^{i2\pi k\theta}
 \qquad \text{and its inverse} \qquad
 q(k) = \int_{\T}\!d\theta\, q^\land(\theta) \mathrm e^{-i2\pi k\theta},
\end{equation}
with $\T$ the torus parametrised by $[0,1)$, the symbol of a
translation-invariant quasi-free state is unitarily equivalent with
the multiplication operator by $q^\land$ on $L^2(\T,d\theta)$. This
function $q^\land$ satisfies $0\le q^\land(\theta)\le 1$ almost
everywhere.

A quasi-free state $\varphi$ is pure if and only if its symbol $Q$ is
a projector. For a translation-invariant state this means that the
Fourier transform of the symbol $Q$ is a characteristic function,
i.e., there exists a measurable set $K\subset\T$ such that
$q^\land(\theta)=\chi_K(\theta)$.

The entropy of a quasi-free state $\varphi$ can be expressed in terms
of its symbol $Q$. Define, for $0\le x\le1$, the functions $\eta(x)
:= -x\log x$ and $\tilde{\eta}(x):= \eta(x) + \eta(1-x)$. The
von~Neumann entropy of the state restricted to an interval of $N$
spins is given by
\begin{equation} \label{eq:entropy}
 S_N := \Tr \eta\left([\varphi_N]\right)
     = \Tr \tilde{\eta}(Q_N),
\end{equation}
where $Q_N$ is the restriction of $Q$ to the $N$-dimensional space
spanned by $\{\delta_k \,:\, 0\leq k\leq N-1\}$. It follows by
Szeg\"o's theorem~\cite{gs} that the entropy density of a
translation-invariant quasi-free state equals
\begin{equation*}
 s := \lim_{N\to\infty} \frac{S_N}{N}
 = \int \!d\theta\, \tilde{\eta}\bigl( q^\land(\theta)\bigr).
\end{equation*}
In particular, the entropy density of a pure translation-invariant
quasi-free state is zero.

\section{Asymptotics for entropy of quasi-free states} \label{sec:asympt}

Quasi-free states are good approximations of true ground and
equilibrium states for systems of Fermions, either at low density or
with weakly interacting particles. The coordinate $\theta$ appearing
in~(\ref{eq:Qcomp}) has the meaning of momentum and the system is
specified by a dispersion relation $\theta\mapsto\varepsilon(\theta)$
which is the relation between effective energy and momentum. For a
shift-invariant quasi-free state, determined by a symbol $Q$ or,
equivalently, by a measurable function $q^\land$ on the unit circle
with $0\leq q^\land\leq 1$, the energy and particle densities are given by
\begin{equation*}
 e(\varepsilon,q^\land) := \int_{\T} \!d\theta\, \varepsilon(\theta)\,
 q^\land(\theta)
 \qquad\text{and}\qquad
 n(q^\land) := \int_{\T} \!d\theta\, q^\land(\theta).
\end{equation*} 
The ground state at density $\lambda$, $0\le\lambda\le1$, is obtained by
minimising the energy density under the constraint $n(q^\land)=\lambda$.
It is given by $q^\land = \chi_{K(e_{\mathrm F}(\lambda))}$, where
$K(e) := \{\theta\in\T\,:\,\varepsilon(\theta)\leq e\}$ and
$e_{\mathrm F}(\lambda)$ is the Fermi level determined by the condition
\begin{equation*}
 \left| K(e_{\mathrm F}(\lambda)) \right| =
 \int_{\varepsilon(\theta)\leq e_{\mathrm F}(\lambda)} \!d\theta =
 \lambda.
\end{equation*} 
For smooth dispersion relations with few oscillations in $\theta$,
$K(e_{\mathrm F}(\lambda))$ will typically consist of a finite union
of disjoint intervals. This case will be investigated
in Section~\ref{sec:finite}. Section~\ref{sec:infinite} deals with
the opposite situation when $K(e_{\mathrm F}(\lambda))$ has
a Cantor-like structure.
 
The quasi-free states on the spin chain $\qsc$, as introduced
in the previous section, obey Equation~(\ref{eq:entropy}) for
the von~Neumann entropy of the restricted density matrices.
This will be the starting point for our study of the asymptotic
behaviour of this entropy $S_N$ as $N\to\infty$.

\subsection{Growth exponents}

\begin{figure}
\begin{center}
\includegraphics[width=8cm]{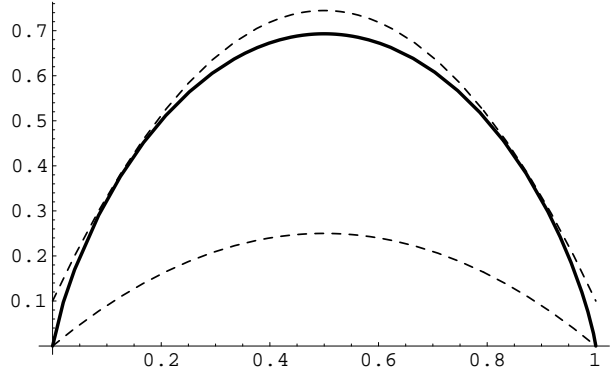}
\end{center}
\caption{\label{fig:bounds} A quadratic upper and lower bound for the entropy 
function $\tilde\eta$.}
\end{figure}

We use the following estimate for the entropy function
$\tilde\eta(x) := \eta(x)+\eta(1-x)$,
\begin{equation*}
 x(1-x) \leq \tilde\eta(x) \leq
 \epsilon - c\, \log\epsilon\ x(1-x),\quad 0\leq x\leq 1,
\end{equation*}
see Figure~\ref{fig:bounds}.
The upper bound for $\tilde\eta$ holds for  $c$ a constant
independent of $0<\epsilon$, moreover, for $0<\epsilon<\epsilon_0$ we
may choose $c=1+\mathrm o(\epsilon_0)$.  Therefore,
\begin{equation*}
 \Tr Q_N(\idty-Q_N) \leq S_N \leq \epsilon N - c\,\log\epsilon\ 
 \Tr Q_N(\idty-Q_N).
\end{equation*}
By choosing a function $\epsilon(N)$ for which $\epsilon\to 0$
as $N\to\infty$, we obtain bounds for the entropy $S_N$ in terms
of $\Tr Q_N(\idty-Q_N)$. E.g., putting $\epsilon(N):=\frac{1}{N}$,
\begin{equation} \label{eq:bound}
\Tr Q_N(\idty-Q_N) \leq S_N \leq
1 + c\,\log N\ \Tr Q_N(\idty-Q_N).
\end{equation}

We are particularly interested in the growth exponents of the entropy,
\begin{equation*}
\alpha_- := \liminf_{N\to+\infty} \frac{\log S_N}{\log N}
\qquad \mbox{and} \qquad
\alpha_+ := \limsup_{N\to+\infty} \frac{\log S_N}{\log N}.
\end{equation*}
With the inequalities~(\ref{eq:bound}),
\begin{equation*}
\liminf_{N\to+\infty} \frac{\log \Tr Q_N(\idty-Q_N)}{\log N}
\leq \alpha_- \leq \alpha_+ \leq
\limsup_{N\to+\infty}  \frac{\log \Tr Q_N(\idty-Q_N)}{\log N}.
\end{equation*}
We conclude that, if
\begin{equation*}
\lim_{N\to+\infty} \frac{\log \Tr Q_N(\idty-Q_N)}{\log N}
\quad \mbox{exists, then also }
\lim_{N\to+\infty} \frac{\log S_N}{\log N}
\quad \mbox{exists, and}
\end{equation*} \begin{equation} \label{eq:exponent}
\alpha := \lim_{N\to +\infty} \frac{\log S_N}{\log N}
= \lim_{N\to +\infty} \frac{\log \Tr Q_N(\idty-Q_N)}{\log N}.
\end{equation}

\subsection{Quadratic approximation}

Equations~(\ref{eq:bound}) and (\ref{eq:exponent}) show the
importance of the quantity $\Tr Q_N(\idty-Q_N)$. It can be expressed
in terms of the sequence $\{q(k)\}$ or, equivalently, of the  Fourier
transform $q^\land(\theta)=\chi_K(\theta)$ of the symbol $Q_N$. Using
Equation~(\ref{eq:Qcomp}),
\begin{align*}
 &\Tr Q_N(\idty-Q_N)  \\
 &\quad= N q(0) - \sum_{n,m=1}^N |q(n-m)|^2 \\
 &\quad= N q(0) - N \sum_{n=-(N-1)}^{N-1}
    \left(1-\frac{|n|}{N}\right) |q(n)|^2 \\
 &\quad= N q(0) - N \int\!d\theta_1 \int\!d\theta_2
 \,\chi_K(\theta_1)\chi_K(\theta_2)
 \sum_{n=-(N-1)}^{N-1} \left(1-\frac{|n|}{N}\right)
 \e^{i2\pi n(\theta_1-\theta_2)}.
\end{align*}
Define the Dirichlet kernel,
\begin{align*}
 k_N(\varphi)
 &:= \sum_{n=-(N-1)}^{N-1}
   \left(1-\frac{|n|}{N}\right) \e^{i2\pi n\varphi} \\
 &= 1 + 2 \sum_{n=1}^{N-1} \frac{N-n}{N} \cos 2\pi n\varphi
 \,=\, \frac{1}{N} \frac{\sin^2 N\pi\varphi}{\sin^2\pi\varphi}.
\end{align*}
This is a sequence of positive normalised functions,
weakly converging to the Dirac distribution, 
\begin{equation*}
 k_N(\varphi) \geq 0 ,\qquad \int\!d\varphi\, k_N(\varphi) = 1.
\end{equation*}
Therefore,
\begin{align}
 \Tr Q_N(\idty-Q_N) 
 &= N\left( \int\!d\theta\, \chi_K(\theta) -\int\!d\theta \int\!d\varphi\, 
 \chi_K(\theta)\chi_K(\theta-\varphi) k_N(\varphi) \right) 
\nonumber \\
 &= N \int\!d\theta \int\!d\varphi\, \chi_K(\theta)
 \left[1-\chi_K(\theta-\varphi)\right] k_N(\varphi) 
\nonumber \\
 &= N \int\!d\varphi\, k_N(\varphi)\int\!d\theta\, \chi_K(\theta)
\left[1-\chi_K(\theta-\varphi)\right] 
\nonumber \\
 &= N \int\!d\varphi\, k_N(\varphi)\,|K\setminus (K+\varphi)|. 
\label{eq:quadr}
\end{align}
Note that both $S(Q_N)$ and $\Tr Q_N(\idty-Q_N)$ are invariant for
the replacement of $Q_N$ by $\idty-Q_N$. As a consequence,
Equation~(\ref{eq:quadr}) can be written in the form
\begin{equation} \label{eq:quadr2}
 \Tr Q_N(\idty-Q_N) = 
 N \int\!d\varphi\, k_N(\varphi)\,|K^c\setminus (K^c+\varphi)|.
\end{equation}

\section{Finitely many intervals} \label{sec:finite}

As explained in Section~\ref{sec:quafree} the subset $K$
of the torus $\T$ determines the state $\varphi$ we are
studying. In this section we study sets $K$ composed
of a finite number of intervals, whereas in the next section
sets with an infinite number of intervals are treated.

\subsection{Lower bound}

By Equation~(\ref{eq:bound}) we have to bound $\Tr Q_N(\idty-Q_N)$
from below. We consider a set $K$ with a finite number of intervals,
say $M$. Let $\delta>0$ be a fixed number which is smaller than
any of the intervals and the holes between two such intervals.
Therefore, $|K\setminus (K+\varphi)|\geq M\varphi$ for
$0\leq\varphi\leq\delta$. Equation~(\ref{eq:quadr}) becomes,
\begin{align*}
 S_N 
 &\ge \Tr Q_N(\idty-Q_N) \\
 &= N \int\!d\varphi\, k_N(\varphi)\,|K\setminus (K+\varphi)| \\
 &\ge 2NM \int_0^{\delta}\!d\varphi\, k_N(\varphi)\,\varphi \\
 &= 2NM \int_0^{\delta}\!d\varphi\,\varphi
    \left[ 1+2\sum_{n=1}^{N-1}\frac{N-n}{N}\cos 2\pi n\varphi \right] \\
 &= NM \left[ \delta^2
      + \frac{2\delta}{\pi}\sum_{n=1}^{N-1} \frac{\sin 2\pi n\delta}{n}
      - \frac{2}{\pi^2}\sum_{n=1}^{N-1}\frac{\sin^2 \pi n\delta}{n^2} \right] \\
 &\quad+ M \left[- \frac{2\delta}{\pi}\sum_{n=1}^{N-1} \sin 2\pi n\delta
      - \frac{1}{\pi^2}\sum_{n=1}^{N-1}\frac{\cos 2\pi n\delta}{n}
      + \frac{1}{\pi^2}\sum_{n=1}^{N-1}\frac{1}{n} \right].
\end{align*}
Using the identities,
\begin{equation*}
 \sum_{n=1}^{+\infty} \frac{\sin 2\pi n\delta}{n}
 = \frac{\pi}{2} (1-2\delta)
 \qquad\text{and}\qquad
 \sum_{n=1}^{+\infty} \frac{\sin^2 \pi n\delta}{n^2}
 = \frac{\pi^2}{2} \delta(1-\delta),
\end{equation*}
we obtain
\begin{align}
 S_N 
 &\ge NM \left[
 - \frac{2\delta}{\pi}\sum_{n=N}^{+\infty} \frac{\sin 2\pi n\delta}{n}
 + \frac{2}{\pi^2}\sum_{n=N}^{+\infty} \frac{\sin^2 \pi n\delta}{n^2}
              \right]
\nonumber \\
 &\quad+ M \left[
 - \frac{2\delta}{\pi}\sum_{n=1}^{N-1} \sin 2\pi n\delta
 - \frac{1}{\pi^2}\sum_{n=1}^{N-1}\frac{\cos 2\pi n\delta}{n}
 + \frac{1}{\pi^2}\sum_{n=1}^{N-1}\frac{1}{n}
      \right]. 
\label{eq:finlow1}
\end{align}

Next, we estimate the different terms in~(\ref{eq:finlow1}). The first 
term on the first line,
\begin{equation*}
 \left| \sum_{n=N}^{+\infty} \frac{\sin 2\pi n\delta}{n} \right|
 = \left| \sum_{n=N}^{+\infty}
   \frac{\cos\pi(2n+1)\delta-\cos\pi(2n-1)\delta}{2n\sin\pi\delta} \right|
 \leq \frac{1}{N} \frac{1}{|\sin \pi\delta|}.
\end{equation*}
The second term on the first line,
\begin{equation*}
 \left| \sum_{n=N}^{+\infty} \frac{\sin^2 \pi n\delta}{n^2} \right|
 \leq \sum_{n=N}^{+\infty} \frac{1}{n^2}
 \leq \int_{N-1}^{+\infty} \!dx\, \frac{1}{x^2}
 = \frac{1}{N-1}.
\end{equation*}
The first term on the second line,
\begin{equation*}
 \left| \sum_{n=1}^{N-1} \sin 2\pi n\delta \right|
 \leq \frac{1}{|\sin \pi\delta|}.
\end{equation*}
The second term on the second line,
for any $\epsilon>0$ and $N$ sufficiently large,
\begin{equation*}
 \left| \sum_{n=1}^{N-1}\frac{\cos 2\pi n\delta}{n} \right|
 \leq -\log \left[ 2\sin(2\pi\delta) \right] + \epsilon.
\end{equation*}
Finally, the last term on the last line,
\begin{equation*}
 \sum_{n=1}^{N-1} \frac{1}{n}
 \ge \int_1^{N} \!dx\, \frac{1}{x} = \log N.
\end{equation*}
Putting everything together in (\ref{eq:finlow1}), we find that
there exists a constant $c_1>0$ independent of $N$ such that
\begin{equation} \label{eq:finlow2}
S_N \geq c_1 \log N.
\end{equation}

\subsection{Subadditivity}

Before establishing the upper bound for the entropy $S_N$
in case the set $K$ is composed of a finite number of intervals,
we prove a general subadditivity property of this entropy.
This will enable us to restrict the proof of the upper bound
to the case of a single interval.

Suppose that $K_1$ and $K_2$ are disjoint sets and put $K:=K_1\cup K_2$.
Denoting the symbols of these states by $Q$, $Q_1$ and $Q_2$,
we shall prove the subadditivity property, namely,
\begin{equation} \label{eq:subadd}
\Tr\tilde{\eta}(Q_N) \leq
\Tr\tilde{\eta}((Q_1)_N) + \Tr\tilde{\eta}((Q_2)_N).
\end{equation}

To simplify notation, define $R:=Q_N$, $R_1:=(Q_1)_N$
and $R_2:=(Q_2)_N$. First, note that $R=R_1+R_2$. Remember
that $\tilde{\eta}(x)=-x\log x-(1-x)\log(1-x)$ and thus 
$\tilde{\eta}'(x)=-\log x+\log(1-x)$. We assume $R_1>0$,
$R_2>0$ and $R_1+R_2<\idty$. Otherwise, we can introduce
operators $\tilde{R}_1:=(1-\epsilon)R_1+\frac{\epsilon}{2}\idty$
and $\tilde{R}_2:=(1-\epsilon)R_2+\frac{\epsilon}{2}\idty$,
prove the subadditivity for these two operators and then take
the limit $\epsilon\to 0$. Using the operator identity
$\frac{d}{d\lambda}\Tr f(A+\lambda B) = \Tr B f'(A+\lambda B)$,
\begin{align}
 \Tr \tilde{\eta}(R_1+R_2) - \Tr \tilde{\eta}(R_1)
 &= \Tr \tilde{\eta}(R_1+\lambda R_2) \bigg|_{\lambda=0}^1
\nonumber \\
 &= \int_0^1\!d\lambda\, \frac{d}{d\lambda} \Tr 
      \tilde{\eta}(R_1+\lambda R_2)
\nonumber \\
 &= \int_0^1\!d\lambda\, \Tr R_2
    \log\frac{\idty-R_1-\lambda R_2}{R_1+\lambda R_2}.
\label{eq:subadd1}
\end{align}
Because the inverse is operator decreasing,
\begin{equation*}
\frac{\idty-R_1-\lambda R_2}{R_1+\lambda R_2}
= \frac{1}{R_1+\lambda R_2}-\idty
\leq \frac{1}{\lambda R_2}-\idty
= \frac{\idty-\lambda R_2}{\lambda R_2},
\end{equation*}
and, because the logarithm is operator increasing,
\begin{equation*}
\log \frac{\idty-R_1-\lambda R_2}{R_1+\lambda R_2}
\leq \log \frac{\idty-\lambda R_2}{\lambda R_2}.
\end{equation*}
Substituting this into Equation~(\ref{eq:subadd1}),
\begin{equation*}
\Tr \tilde{\eta}(R_1+R_2) - \Tr \tilde{\eta}(R_1)
\leq \Tr \tilde{\eta}(R_2).
\end{equation*}

\subsection{Upper bound}

Due to subadditivity, it is enough to prove an upper bound
for a set $K$ consisting of a single interval. We assume
that the length of this interval $|K|\leq \frac{1}{2}$.
Otherwise, we can work with $K^c$. By Equation~(\ref{eq:bound})
we have to bound $\Tr Q_N(\idty-Q_N)$. In this case,
\begin{equation*}
|K\setminus (K+\varphi)|
= \begin{cases} \varphi &\text{for } |\varphi|\le|K|, \\
  |K| &\text{for } |K|\le|\varphi|\le\frac{1}{2}.
\end{cases}
\end{equation*}
By Equation~(\ref{eq:quadr}),
\begin{align*}
 \Tr Q_N(\idty-Q_N)
 &= 2N \int_0^{\frac{1}{2}} \!d\varphi\,
    \left[ 1+2\sum_{n=1}^{N-1}\frac{N-n}{N}\cos 2\pi n\varphi \right]
    |K\setminus (K+\varphi)| \\
 &= N \left[ |K| (1-|K|) -
     \frac{2}{\pi^2} \sum_{n=1}^{N-1} \frac{\sin^2 \pi n|K|}{n^2}
      \right] \\
 &\quad + \frac{2}{\pi^2} \sum_{n=1}^{N-1} \frac{\sin^2 \pi n|K|}{n}.
\end{align*}
Using the identity,
\begin{equation*}
 \sum_{n=1}^{+\infty} \frac{\sin^2 \pi n|K|}{n^2}
 = \frac{\pi^2}{2} |K| (1-|K|),
\end{equation*}
we obtain,
\begin{equation} \label{eq:finupp1}
 \Tr Q_N(\idty-Q_N)
 = \frac{2N}{\pi^2} \sum_{n=N}^{+\infty} \frac{\sin^2 \pi n|K|}{n^2}
 + \frac{2}{\pi^2} \sum_{n=1}^{N-1} \frac{\sin^2 \pi n|K|}{n}.
\end{equation}
The first term,
\begin{equation*}
 \sum_{n=N}^{+\infty} \frac{\sin^2 \pi n|K|}{n^2}
 \leq \sum_{n=N}^{+\infty} \frac{1}{n^2}
 \leq \int_N^{+\infty} \!dx\, \frac{1}{x^2} = \frac{1}{N}.
\end{equation*}
The second term,
\begin{equation*}
 \sum_{n=1}^{N-1} \frac{\sin^2 \pi n|K|}{n}
 \leq \sum_{n=1}^{N-1} \frac{1}{n}
 \leq 1+\int_1^{N-1} \!dx\, \frac{1}{x} = 1+\log(N-1).
\end{equation*}
Putting everything together in (\ref{eq:finupp1}), we find that
there exists a constant $c_2$ independent of $N$ such that
\begin{equation*}
\Tr Q_N(\idty-Q_N) \leq c_2 \log N,
\end{equation*}
and, finally, by Equation~(\ref{eq:bound}), there exists
a constant $c_3$ independent of $N$ such that
\begin{equation} \label{eq:finupp2}
S_N \leq c_3 \, (\log N)^2.
\end{equation}

\subsection{Numerical results}

\begin{figure}
\begin{center}
\includegraphics[width=8cm]{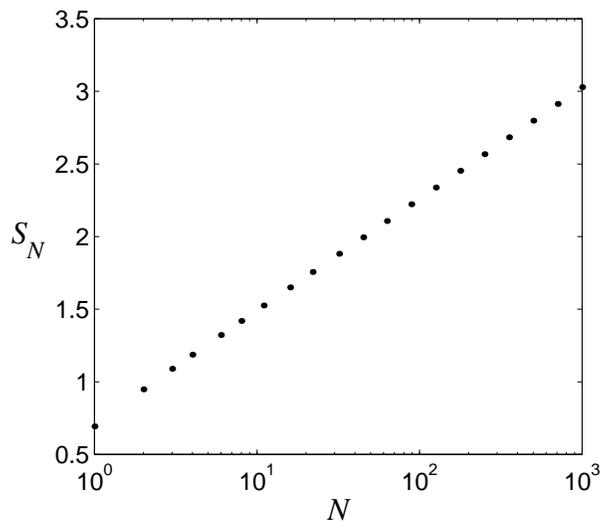}
\end{center}
\caption{\label{fig:logN} The entropy $S_N$ as a function of
the length $N$ of the restriction, for an interval of length
$\frac{1}{2}$ as set $K$. Notice the logarithmic scale.}
\end{figure}

Analytically, we determined the asymptotics of the entropy $S_N$
between the $\log N$ lower bound (\ref{eq:finlow2})
and the $(\log N)^2$ upper bound (\ref{eq:finupp2}).
In Figure~\ref{fig:logN} we present the results of a numerical
calculation. The set $K$ consists of one interval of length
$|K|=\frac{1}{2}$. The figure shows clearly the $\log N$
dependence. By the subadditivity property~(\ref{eq:subadd}),
we expect the same behaviour for all sets $K$ consisting of
a finite number of intervals.

\section{Infinitely many intervals} \label{sec:infinite}

For a set $K$ consisting of finitely many intervals, the entropy
$S_N$ increases asymptotically slower than any power $N^\alpha$ with
$\alpha>0$. However, it is easy to construct a state such that $S_N =
N \log 2$. For example, one can take for $K$ a set of $2^{N-1}$
regularly spaced intervals, each of length $2^{-N}$. Note that this
construction does not have an appropriate limit when $N\to\infty$.
Nevertheless, it suggests that in the infinitely many intervals case
the entropy $S_N$ could have a richer behaviour. This will be shown
in the present section by presenting a class of states for which the
growth exponent $\alpha$ can take any value $\alpha\in(0,1)$.

\subsection{A Cantor-like construction}

The standard Cantor set is constructed recursively by removing in
step $m$ a fixed fraction of the set obtained in step $m-1$. This
would leave us with a set of zero Lebesgue measure. To avoid this, we
remove a fraction in step $m$ which decreases with $m$, such that
the limit set has strictly positive Lebesgue measure.

We start with the unit interval. Remove in the first step an open
interval in the middle of the unit interval with length
$1-\gamma(1)$. The resulting set $K_1$ consists of two closed
intervals each of length $\gamma(1)/2$. In the second step, for each
of these two intervals, a fraction $1-\gamma(2)$ is removed in the
middle of these intervals. This leaves us with a set $K_2$ consisting
of four closed intervals of length $\gamma(1)\gamma(2)/4$. Continuing
this procedure, in step $m$ we obtain a set $K_m$ of $2^m$ closed
intervals of length
\begin{equation} \label{eq:intervals}
 \ell_i(m) := \frac{\prod_{n=1}^{m}\gamma(n)}{2^m}.
\end{equation}
There are $2^m-1$ holes in between such intervals, $2^{m-1}$ of which
are created in step $m$. The latter have a length
\begin{equation} \label{eq:holes}
 \ell_h(m) := \ell_i(m-1)(1-\gamma(m))
 = \frac{\prod_{n=1}^{m-1}\gamma(n)}{2^{m-1}}(1-\gamma(m)).
\end{equation}
The Lebesgue measure of the limit set $K$ is then
$\prod_{n=1}^\infty\gamma(n)$.

To construct an explicit example, we have to fix a function
$m\mapsto\gamma(m)$. We can as well specify the function
$m\mapsto \ell_h(m)=aq^m$, where $0<q<\frac{1}{2}$
and $a$ is chosen such that
\begin{equation*}
 1 > \sum_{m=1}^{+\infty} 2^{m-1} aq^m = \frac{a}{2}\,  \frac{2q}{1-2q}.
\end{equation*} 
The resulting set has Lebesgue measure $1-aq/(1-2q)$.

\subsection{Lower bound}

To bound $\Tr Q_N(\idty-Q_N)$ from below, we start from
Equation~(\ref{eq:quadr}). As before, $K_m$ denotes the set obtained
after $m$ steps in the construction of the Cantor-like set $K$.
Then $K_m$ is the union of $2^m$ intervals, each of length $l_i(m)$.
Because $K\subset K_m$, (\ref{eq:quadr}) can be estimated by
\begin{align*}
 \Tr Q_N(\idty-Q_N)
 &=    N \int_{-\frac{1}{2}}^{\frac{1}{2}}\!d\varphi\,
         k_N(\varphi) |K\setminus (K+\varphi)| \\
 &\ge N \int_{-\frac{1}{2}}^{\frac{1}{2}}\!d\varphi\,
         k_N(\varphi) |K\setminus (K_m+\varphi)|,
\end{align*}
and since $k_N(\varphi)\ge N/\pi^2$ when $|\varphi|\le1/2N$,
\begin{align*}
 \Tr Q_N(\idty-Q_N)
 &\ge \frac{1}{\pi^2}N^2 \int_{-\frac{1}{2N}}^{\frac{1}{2N}} \!d\varphi\,  
         |K\setminus (K_m+\varphi)| \\
 &= \frac{2}{\pi^2}N^2 \int_{0}^{\frac{1}{2N}} \!d\varphi\,
         |K\setminus (K_m+\varphi)|.
\end{align*}

For given $N$, take $m$ such that
\begin{equation} \label{eq:lowerNm}
 \ell_h(m)\ge\frac{1}{2N}>\ell_h(m+1).
\end{equation}
As $K_m$ consists of $2^m$ translations of the interval $[0,\ell_i(m)]$,
the Cantor-like set $K$ consists of $2^m$ translations of another
Cantor-like set $\tilde{K}\subset[0,\ell_i(m)]$. Let us denote
these translations by $x_\ell+[0,\ell_i(m)]$ and $x_\ell+\tilde{K}$ for
$\ell=1,\ldots,2^m$.

For $\varphi\in [0,1/2N]$, $\varphi\leq \ell_h(k)$ for all
$k=1,\ldots,m$. This means that a translation by $\varphi$ of an
interval of length $\ell_i(m)$ in $K_m$ will never bridge the hole
(of length $\ell_h(k), k=1,\ldots,m$) between this interval and the
next. Therefore, every $x_\ell+\tilde{K}$ will overlap with one and
only one $x_{\tilde{\ell}}+[0,\ell_i(m)]+\varphi$, namely the one
with $\tilde{\ell}=\ell$. As a consequence,
\begin{equation*}
 \Big| K\setminus (K_m+\varphi) \Big|
 = 2^m \Big| \tilde{K}\setminus ([0,\ell_i(m)]+\varphi) \Big|
 = 2^m \Big| \tilde{K}\setminus [\varphi,\ell_i(m)] \Big|
 = 2^m \Big| \tilde{K} \cap [0,\varphi] \Big|.
\end{equation*}

This quantity has to be estimated from below.
For $\varphi\in (0,1/2N]$, take $n$ such that
$\ell_i(n)\geq\varphi>\ell_i(n+1)$. Then,
\begin{equation*}
 \left| \tilde{K}\cap [0,\varphi] \right|
 \geq \ell_i(n+1)\prod_{k=n+2}^{+\infty}\gamma(k)
 = \frac{1}{2^{n+1}}\prod_{k=1}^{+\infty}\gamma(k),
\end{equation*}
and
\begin{equation*}
\varphi\leq \ell_i(n)= \frac{1}{2^n}\prod_{k=1}^{n}\gamma(k),
\end{equation*}
which gives
\begin{equation*}
\frac{\left| \tilde{K}\cap [0,\varphi] \right|}{\varphi}
\geq \frac{1}{2}\prod_{k=n+1}^{+\infty}\gamma(k)
\geq \frac{1}{2}|K|,
\end{equation*}
and so
\begin{equation*}
\left| \tilde{K}\cap [0,\varphi] \right| \geq \frac{1}{2}|K|\varphi,
\end{equation*}
which does not depend any longer on $n$. It follows that
\begin{equation*}
\Tr Q_N(\idty-Q_N)
\geq 2^m \frac{1}{\pi^2}|K| N^2 \int_{0}^{\frac{1}{2N}}\!d\varphi\,\varphi
=    2^m \frac{1}{8\pi^2}|K|.
\end{equation*}

This is an estimate from below of $\Tr Q_N(\idty-Q_N)$
where $N$ and $m$ are coupled by (\ref{eq:lowerNm}).
From the latter we also have that $N<1/2\ell_h(m+1)$.
Therefore,
\begin{equation*}
\frac{\log \Tr Q_N(I-Q_N)}{\log N} > \frac{\log\left(2^m\frac{1}{8\pi^2}|K|\right)}
{-\log\left(2\ell_h(m+1)\right)}.
\end{equation*}
The limit $N\to\infty$ corresponds to the limit $m\to\infty$.
Using the explicit form $\ell_h(m)=aq^m$, we finally get
\begin{equation} \label{eq:liminf}
\liminf_{N\to+\infty} \frac{\log \Tr Q_N(I-Q_N)}{\log N}
\geq \frac{\log 2}{-\log q}.
\end{equation}

\subsection{Upper bound}

To get an upper bound for $\Tr Q_N(\idty-Q_N)$, we start from
Equation~(\ref{eq:quadr2}).
With $C(\varphi):=|K^c\setminus (K^c+\varphi)|$, it reads
\begin{equation*}
\Tr Q_N(\idty-Q_N) = N \int\!d\varphi\, k_N(\varphi)C(\varphi),
\end{equation*}

For $\theta>0$, take $m$ such that
\begin{equation} \label{eq:upperthm}
\ell_h(m) \geq \theta > \ell_h(m+1).
\end{equation}
We bound $C(\theta)$ from above,
\begin{align}
 C(\theta) 
 &\le \sum_{k=1}^\infty 2^{k-1} \min\{\theta,\ell_h(k)\} 
\nonumber \\
 &= \sum_{k=1}^{m} 2^{k-1} \theta + \sum_{k=m+1}^\infty 2^{k-1}
 \ell_h(k) 
\nonumber \\
 &\le (2^m-1) \ell_h(m) + \sum_{k=m+1}^\infty 2^{k-1} \ell_h(k) 
\nonumber \\
 &\le 2 \prod_{n=1}^{m-1} \gamma(n)\,(1-\gamma(m)) +
 \sum_{k=m+1}^\infty \prod_{n=1}^{k-1} \gamma(n)\, (1-\gamma(k)). 
\label{eq:upper1}
\end{align}
Obviously, this bound increases with $\theta$.
The kernel $k_N(\varphi)$ satisfies
\begin{equation*}
 k_N(\varphi) \le 
 \begin{cases} N &\text{for } |\varphi|\le\theta, \\
 \frac{\pi^2}{2N}\, \frac{1}{\varphi^2} &\text{for } |\varphi|\ge\theta,
 \end{cases}
\end{equation*}
and so we find
\begin{align}
 \Tr Q_N(\idty-Q_N) 
 &\le N^2 \int_{|\varphi|\le\theta} \!d\varphi\, C(\varphi) +
 \int_{|\varphi|\ge\theta} \!d\varphi\, C(\varphi) \frac{\pi^2}{2}\,
 \frac{1}{\varphi^2} 
\nonumber \\
 &\le 2N^2 \theta C(\theta)
 + \frac{\pi^2}{2} \sum_{k=0}^{m}
 (\ell_h(k)-\ell_h(k+1)) \frac{C(\ell_h(k))}{\ell_h(k+1)^2} 
\nonumber \\
 &\le 2N^2 \ell_h(m) C(\theta)
 + \frac{\pi^2}{2} \sum_{k=0}^{m}
 \frac{\ell_h(k)C(\ell_h(k))}{\ell_h(k+1)^2}. 
\label{eq:upper2}
\end{align}

Take again the explicit form $\ell_h(m)=aq^m$. Then
\begin{equation*}
 \prod_{n=1}^{m-1} \gamma(n)\, (1-\gamma(m)) = 2^{m-1}\ell_h(m) =
 \frac{a}{2} (2q)^m,
\end{equation*}
and (\ref{eq:upper1}) becomes
\begin{align}
 C(\theta) 
 &\le 2\prod_{n=1}^{m-1} \gamma(n)\, (1-\gamma(m)) +
 \sum_{k=m+1}^\infty \prod_{n=1}^{k-1} \gamma(n)\, (1-\gamma(k))
\nonumber \\
 &= 2\frac{a}{2} (2q)^m + \sum_{k=m+1}^\infty \frac{a}{2} (2q)^{k} =
 \frac{a(1-q)}{1-2q} (2q)^m. 
\label{eq:upperCth}
\end{align}
If $\theta=\ell_h(k)$, then by (\ref{eq:upperthm}), we have to put $m=k$,
and so
\begin{equation} 
\label{eq:upperCk}
 C(\ell_h(k)) \le \frac{a(1-q)}{1-2q} (2q)^k.
\end{equation}
Substituting inequalities~(\ref{eq:upperCth}) and~(\ref{eq:upperCk})
into~(\ref{eq:upper2}), we find
\begin{align}
 \Tr Q_N(\idty-Q_N) 
 &\le 2N^2aq^m \frac{a(1-q)}{1-2q} (2q)^m + \frac{\pi^2}{2}
 \sum_{k=0}^m \frac{aq^k}{(aq^{k+1})^2} \frac{a(1-q)}{1-2q} (2q)^k 
\nonumber \\
 &= 2N^2aq^m \frac{a(1-q)}{1-2q} (2q)^m + \frac{\pi^2}{2}
 \frac{1}{q^2} \frac{2(1-q)}{1-2q} (2^{m+1}-1)
\nonumber \\ 
 &\le \frac{2(1-q)}{1-2q} \left[ N^2a^2(2q^2)^m + \frac{\pi^2}{2}
 \frac{1}{q^2} 2^m \right]
\nonumber \\
 &=: c_1 N^2 (2q^2)^m + c_2 2^m, 
\label{eq:upper3}
\end{align}
where $c_1$ and $c_2$ are independent of $N$.

To get an upper bound as a function of $N$, we have to
fix a function $m(N)$ and plug it into (\ref{eq:upper3}).
Let 
\begin{equation*}
 \gamma := \frac{\log 2}{-\log q} = \frac{1}{-\log_2 q},
\end{equation*} 
then choose $m$ to be
\begin{equation*}
 m = \left[ \log_{2q^2}N^{\gamma-2} \right] \le \log_{2q^2}N^{\gamma-2}
 = \frac{\log_2 N^{\gamma-2}}{\log_2 2q^2},
\end{equation*}
where $[a]$ denotes the integer part of the number $a$.
Then
\begin{equation*}
 N^2 (2q^2)^m \le N^{\gamma},
\end{equation*}
and
\begin{equation*}
 2^m \le \left(2^{\log_2 N^{\gamma-2}}\right)^{\frac{1}{\log_2 2q^2}}
 = N^{\frac{\gamma-2}{1+\log_2 q^2}} = N^{\gamma},
\end{equation*}
and so
\begin{equation*}
\Tr Q_N(\idty-Q_N) \leq (c_1+c_2)N^{\gamma},
\end{equation*}
from which we get the upper bound
\begin{equation} \label{eq:limsup}
 \limsup_{N\to+\infty} \frac{\log \Tr Q_N(\idty-Q_N)}{\log N}
 \le \gamma = \frac{\log 2}{-\log q}.
\end{equation}

Combining the results~(\ref{eq:liminf}) and~(\ref{eq:limsup}) we see
that $\lim_{N\to+\infty} \log \Tr Q_N(\idty-Q_N)/\log N$ exists,
which implies that also $\lim_{N\to+\infty} \log S_N/\log N$ exists,
and
\begin{equation*}
 \alpha = \lim_{N\to+\infty} \frac{\log S_N}{\log N}
 = \lim_{N\to+\infty} \frac{\log \Tr Q_N(\idty-Q_N)}{\log N}
 = \frac{\log 2}{-\log q}.
\end{equation*}
Since $q$ can be any number in the interval $(0,1/2)$,
the growth exponent $\alpha$ can take any value in $(0,1)$.

\end{document}